Chapter XX

# Time-Resolved Optical Studies of Quasiparticle Dynamics in High-Temperature Superconductors

## Experiments and theory

D. Mihailovic and J. Demsar
Jozef Stefan Institute, Jamova 39, 1001 Ljubljana, Slovenia

Ultrafast time-resolved optical spectroscopy in high-temperature superconductors enables the direct real-time measurement of non-equilibrium quasiparticle recombination dynamics. In addition, it also gives detailed information about the symmetry of the superconducting gap and the "pseudogap", their doping dependence and their temperature dependence. Experimental data, together with theoretical models on the photoinduced transmission amplitude and relaxation time as a function of temperature and doping in $YBa_2Cu_3O_{7-\delta}$ gives a consistent picture of the evolution of low-energy structure, where a temperature-independent gap is shown to exist in the underdoped state which evolves with doping into a two-component state near optimum doping, where the dominant response is from a T-dependent BCS-like superconducting gap.

**Introduction**

The experimental study of non-equilibrium phenomena with real-time techniques has been used quite effectively in the past to investigate the quasiparticle (QP) relaxation dynamics in "conventional" non-cuprate superconductors (*1*). The main interest in the QP dynamics has been in superconducting junction devices, where the relaxation of carriers injected into a superconductor has been measured. The typical relaxation times in conventional superconductors are in the range $10^{-7}$ - $10^{-9}$s and the quasiparticle relaxation processes could be experimentally measured by using fast electronics (*2*). In high temperature superconducting cuprates however, because $T_c$ is higher and the superconducting gap is larger, the timescales are much shorter and electrical measurements become very difficult if not impossible with current state of the art electronics.

The availability of reliable ultrafast laser systems in the last decade has made possible the use of ultrafast optical investigations of QP relaxation and recombination phenomena with real-time resolution exceeding 20 fs. A number of time-resolved optical studies since the discovery of these materials have shown that potentially the technique could be a powerful new tool for the study of high-temperature superconductors (*3-8*). In particular, the separation of timescales enables quasiparticle relaxation processes to be distinguished from localized (possibly intra-gap) states (*7*). Rapid progress has been made particularly in the last few years, partly because of systematic measurements as a function of temperature and doping, and partly because of the development of theoretical models to describe the temperature dependence of the transient changes in optical constants and relaxation time in a superconductor (*9*). In this chapter we will describe the current state of the experiments and theory of ultrafast time-resolved laser spectroscopy as applied to high-temperature superconductors (HTS).

Optical spectroscopy, including infrared reflectivity and Raman spectroscopy has been very important in elucidating the low-frequency electronic excitation spectra of high temperature cuprate superconductors (*10-13*). However, because of the large number of ions in the unit cell the low-energy electronic structure becomes complicated, and it becomes difficult to distinguish between different overlapping spectral contributions. Because accurate identification of these different contributions - for example the Drude contribution from conduction electrons or the so-called mid-infrared contribution thought to come from polaron hopping - is crucial for the development of a theory of superconductivity, there has been a significant amount of controversy regarding the interpretation of the low-frequency spectra. The information provided by study of the QP recombination dynamics enables the identification of these different contributions on the basis of their lifetimes as well as the temperature dependences and doping systematics of photoinduced transmissivity or reflectivity data.

At this time there is an additional motivation for the study of non-equilibrium phenomena in these materials, because time-resolved experiments can give information about the gap structure of the superconductor, and consequently also about the nature of the carriers involved in superconductivity and consequently also about the pairing mechanism itself. For example, trapped, localized or polaronic states and correlated electrons are expected to have very different dynamics than charge carriers in extended band states.

Apart from the new information about the low-energy electronic structure, the study of non-equilibrium carrier dynamics gives us a thorough understanding of carrier relaxation dynamics, which is crucial also for the design and construction of non-equilibrium superconducting devices. Since the relaxation time in a superconductor essentially scales inversely with the magnitude of the superconducting gap (*9*), high temperature superconducting cuprates are likely to have much faster relaxation timescales than conventional superconductors, a consideration which is clearly important for high-speed electronics.

                                                                                              3

## The Experimental Technique and Data Analysis

Time-resolved optical spectroscopy of superconductors involves measurement of the transient change of the optical transmission $T$ through, or reflection $R$ from a superconductor by a two step process. In the first step, the superconductor is excited by an ultrashort "pump" laser pulse and in the second, the change $DR/R$ or $DT/T$ of the weaker "probe" pulse is measured as a function of time delay after the photoexcitation. A typical experimental setup is shown schematically in Figure 1. A mode-locked laser (usually Ti:sapphire) is used to generate the laser pulses with pulselength $\tau_p$ ranging from 50-200 fs at around 800 nm (1.5 eV). The pulse energies are typically in the range 0.1-1 nJ.

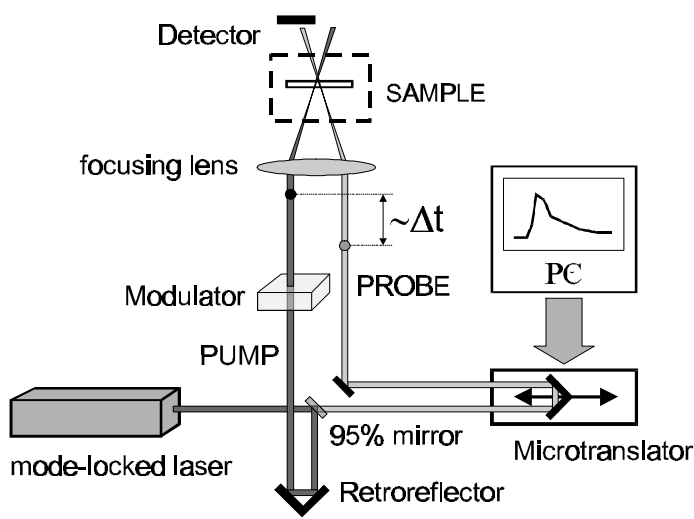

Figure 1. A typical time-resolved photoinduced transmission setup. A similar setup is also used for measuring the photoinduced reflection, except that the reflected probe beam is detected in this case.

The carrier density excited by the probe pulse is typically small compared to the normal state hole density in the cuprates, so the perturbation caused by the photoexcited carriers can be assumed to be small. The probe beam, which is usually (but not always) at the same wavelength as the pump beam, is suitably delayed in time with a Michelson interferometer and significantly attenuated, typically by a factor of 100. Usually the effect of additional photoexcitation by the probe beam can be neglected. Other laser systems have also been used in the past (*14,15*), for example the colliding-pulse mode-locked (CPM) dye laser, and amplified lasers, in the latter case the pump energies may be up to three orders of magnitude higher and can be as high as 1µJ. In this case the photoexcited carrier density can exceed the normal state carrier density.

The changes $DR/R$ or $DT/T$ in the case of weak photoexcitation are small ($10^{-6}$-$10^{-3}$), so high-frequency lock-in detection at $\nu \sim 100$ kHz or more is usually used to reduce laser noise. It is usually sufficient to modulate only the pump beam and detect the probe light with a lock-in detector, but dual modulation techniques are sometimes necessary when bad sample surface quality prevents pump scattered light from being sufficiently rejected.

The typical time-resolved photoinduced transmission signal on near optimally doped $YBa_2Cu_3O_{7-\delta}$ in the range from − 6 ps to 13 ps (open squares) together with the fit (solid line) is presented in Figure 2. The signal is zero when the pump pulse is blocked (*A*), however there is a non-zero long lived component (*7,16*) present even at 12 ns after the pump pulse, seen as a signal at negative times, when the pump is unblocked (*B*). Besides the very slow component there's also a fast transient present, with a rise time of approximately 300 fs and a decay time of approximately 2 ps. From the temperature dependence of the time-resolved photoinduced transmission (reflection) spectra one can obtain the temperature dependence of the fast and slow component amplitude, indicated approximately by a difference between points *C* and *D* and points *D* and *B* respectively, together with the temperature dependence of typical relaxation time.

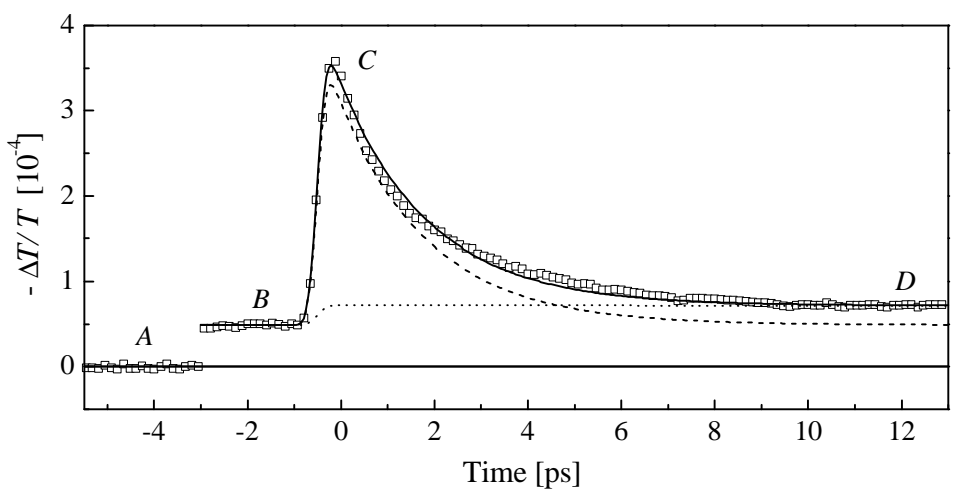

Figure 2: Typical shape of the time-resolved photoinduced transmission signal taken at the cryostat temperature 45 K on near optimally doped $YBa_2Cu_3O_{7-\delta}$ with $T_c$ = 90 K (squares) using 200 fs Ti:sapphire laser pulses, together with the fit using a sum of the slow (dotted line) and fast (dashed line) component, with the effective rise time $\sigma \sim$ 300 fs.

In the low repetition experiments using CPM laser, the slow component was found to be present even after 160 μs (*16*), therefore in the range of several 10 ps it can be considered as a constant and fitted by the Heaviside step function (zero at negative time delays and constant at t > 0). The fast component, on the other hand, was found to be reasonably well reproduced by single exponential decay with a characteristic relaxation time in the range 0.5 – 3 ps depending on doping and temperature (*8,9*). The fit used in the Figure 2 is therefore a sum of the two (fast and slow) components.

Since the rise time of the transient signal is pulse-width limited, in the case of single exponential decay the fast component of photoinduced transmission due to pump pulse is a solution of

$$\partial n / \partial t = -n/t_R + g(t); \qquad n(t) \equiv \Delta T/T(t), \qquad (1)$$

$t_R$ being the relaxation time, and $g(t) = A\exp(-2t^2/t_p^2)$ representing the photoexcitation with pump pulse width $t_p$. Due to finite probe pulse width the resulting $n(t)$ should be convoluted with $g(t')$ to obtain the measured $n(t)$, however since it effects the rise time of the signal only, $n(t)$ can be approximated with the solution of the equation 1 with an *effective* pulse width $t_p' \approx \sqrt{2}t_p$

$$n(t) = C\exp[-t/t_R]\{1 - Erf[(-4tt_R + t_p'^2)/(2\sqrt{2}t_p' t_R)]\} \qquad . \qquad (2)$$

As can be seen in Figure 2, the signal can be very well reproduced using a sum of a Heaviside step function and a single exponential decay both having a pulse width limited risetime.

A typical set of time-resolved photoinduced transmission traces in an underdoped $YBa_2Cu_3O_{7-\delta}$ sample with a $T_c = 53$ K and an optimally doped sample with $T_c = 90$ K obtained with a 200 fs Ti:sapphire setup are shown in Figure 3.

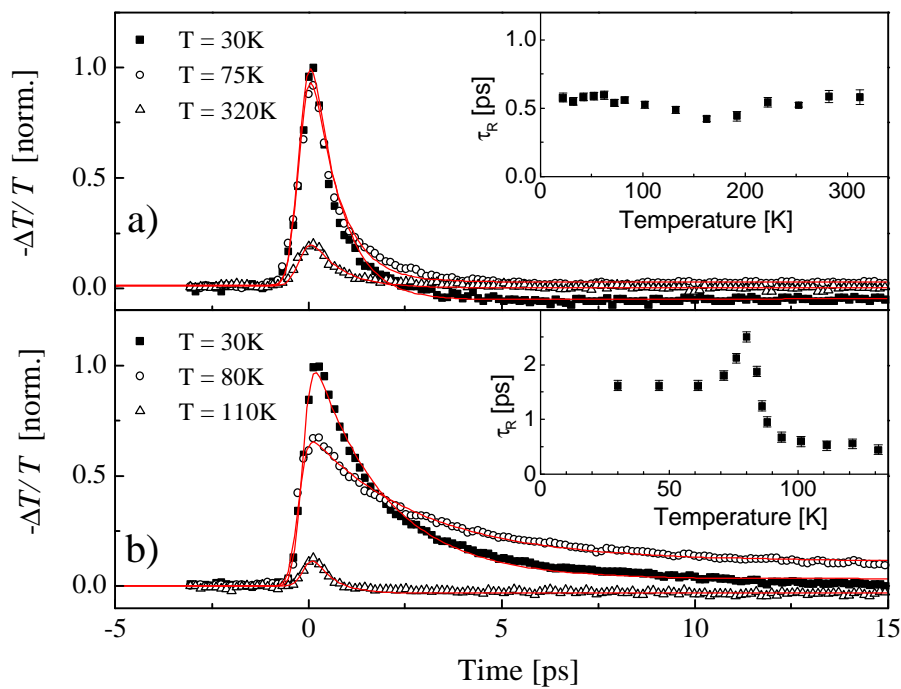

Figure 3: Typical set of time-resolved photoinduced transmission traces in an a) underdoped ($\delta = 0.44$, $T_c = 53K$) and b) near optimally doped ($\delta = 0.08$, $T_c = 90K$) $YBa_2Cu_3O_{7-\delta}$ thin film on SrTiO substrate. The straight lines are fits to the data using single exponential decay with $\tau_R$ in the range 0.5 – 3ps and a Heaviside step function representing the long-lived component (*7,16*). In the insets there are the temperature dependences of the relaxation time obtained using single exponential fit showing completely different behaviors in underdoped and near optimally doped state.

The data for a number of temperatures clearly show that both the lifetime of the decay and the amplitude of the signal are dependent on doping and temperature. A much longer decay time is seen below $T_c$ in the sample with $T_c = 90$ K (inset to Figure 3b)), and a divergence below $T_c$, which is conspicuously absent in the underdoped sample (inset to Figure 3a)). The amplitude, on the other hand, increases with decreasing temperature in both cases.

**Heating Effects.** Laser heating of the sample is an important experimental problem in these experiments and has often been discussed in the literature (*16-18*). There are two effects to be considered. Firstly and more importantly the steady-state temperature of the probed volume increases due to the pump laser excitation. Since the thickness of the substrate or single crystal (~0.3 mm) is typically much bigger than the absorption length (~ 80 nm) one can calculate the temperature rise using a simple steady-state heat diffusion model (*17*), where the Gaussian laser beam with the average laser power $P_L$ is focused into a spot of diameter $d$ on a semi-infinite solid with reflectivity $R$, absorption coefficient $a$, and thermal conductivities $\kappa_x$, $\kappa_y$, $\kappa_z$. Choosing the center of the beam at the crystal surface as the origin of the coordinate system (z=0 at the surface and z>0 in the crystal), the boundary condition is that the temperature at $z = \infty$ is equal to the temperature of the cold finger. Since we are dealing with temperature rises of the order of 10 K, we can neglect the energy loss due to thermal radiation. Using this model we obtain the expression for the temperature rise

$$\Delta T(x,y,z) = \frac{a(1-R)P_L^3 e^{-az}}{4\pi^2 d^2 k_x k_y k_z} \times$$

$$\int_0^{2\pi} d\varphi \int_0^\infty r \cdot dr \exp[-(\tfrac{x}{d} - \tfrac{r\cos\varphi}{k_x d})^2 - (\tfrac{y}{d} - \tfrac{r\sin\varphi}{k_y d})^2][\int_{k_z z}^\infty x(u)du + \int_{-k_z z}^\infty x(u)du] \quad (3)$$

with $x(u) = e^{-au/k_z}/\sqrt{r^2 + u^2}$, $k_x^2 = k_y k_z / P_L^2$ and $k_y$ and $k_z$ respectively and $\varphi$, $r$ and $u$ are the integration variables. In case of single crystals at 80K using $a = 1.8 \times 10^7$ m$^{-1}$ (*19*), $k_x = k_y = 8$ W/mK, $k_z = 2$ W/mK (*20*), $R$~0.1, $d = 100$ μm and $P_L = 10$ mW one obtains $\Delta T(0,0,0) = 12$ K. In the case of experiments on thin films, the heat flow is determined by thermal properties of the substrate, since the film thickness is negligible in comparison with the substrate thickness. Therefore, using the value $k_x = k_y = k_z \sim 18$ W/mK of SrTiO$_3$ at 80 K (*21*) one obtains $\Delta T(0,0,0) = 2.7$ K.

Experimentally, CW heating can also be measured and accounted for quite accurately and is in agreement with the calculation. On thin film samples one can measure the effect of the laser heating with a four-point probe resistivity measurement of a thin microbridge made of the same superconductor film (Figure 4).
Although this method works very well with thin films it is not applicable to single crystals. However, since the temperature dependence of the fast component in near optimally doped superconductors is strongly temperature dependent as shown in Figure 5, and it has been experimentally shown (*9*) that the amplitude at low excitations is linearly dependent on pump fluence, one can often obtain ΔT quite accurately by scaling the temperature dependence of the fast component by normalizing the signal amplitude to the pump fluence and by adjusting the temperature scale due to heating.

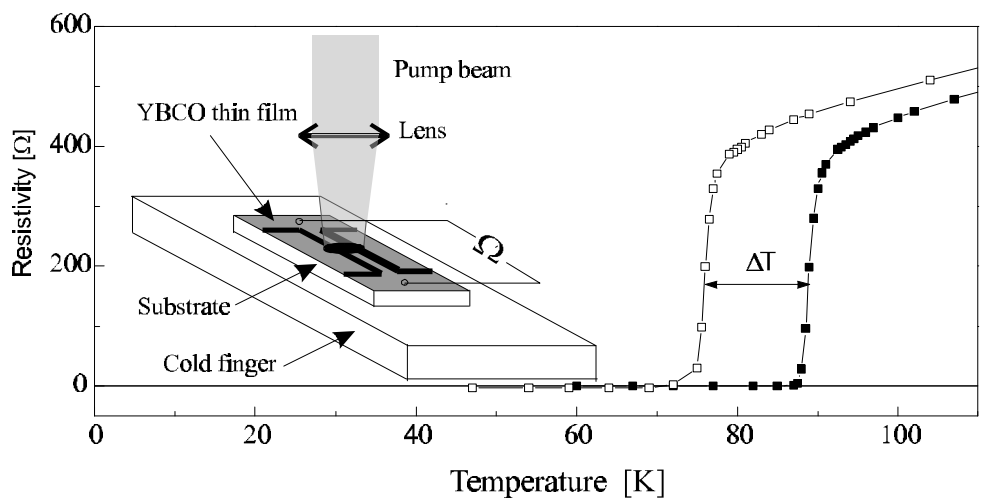

Figure 4: Experimental determination of $\Delta T$ at temperature close to $T_C$ by measuring R(T) of the illuminated microbridge (open squares) compared to the non-illuminated (solid squares). At $P_L = 40$mW, $d = 100$ µm one obtains $\Delta T_{(80K)} \sim 13$ K, which is consistent with the calculated result using equation 2 giving $\Delta T_{(80K)} \sim 11$ K.

In figure 5 there is a temperature dependence of the photoinduced transmission amplitude taken on the same thin film as used in figure 4 at three different laser fluences, where $\mathbf{D}T = 15$ K at $P_L = 40$ mW was determined, again comparable with the calculated temperature rise. When conducting the same analysis on single crystals using the temperature dependence of photoinduced reflectivity amplitude one obtains the result $\mathbf{D}T = 16 - 22$K at $P_L = 10$ mW, which is doping dependent.

Comparison with calculations made with a simple steady-state heat diffusion model give good quantitative agreement and as a result one can determine the temperature of the sample to within +/-2K in the temperature range 4 - 300K. It is important to note that because the thermal constants are very different for different superconductors (and substrates), different experimental configurations can give rise to heating effects which vary by more than one order of magnitude.

The second type of heating effect is non-equilibrium heating, where the temperature builds up during and after the pump laser excitation but dissipates before the next laser pulse. This effect can also be calculated accurately using the heat-diffusion model (*16,18*) with transient laser pulses including the anisotropy in thermal conductivity. The effect is smaller however, and typical values of the peak transient temperature increases in the low excitation experiments are *0.1-0.3* K/mW, so for most common experimental configurations it can be ignored. The effect is much more important if more intense laser pulses are used, from an amplified laser for example. In such cases, the non-equilibrium heating effects need to be carefully considered in the analysis.

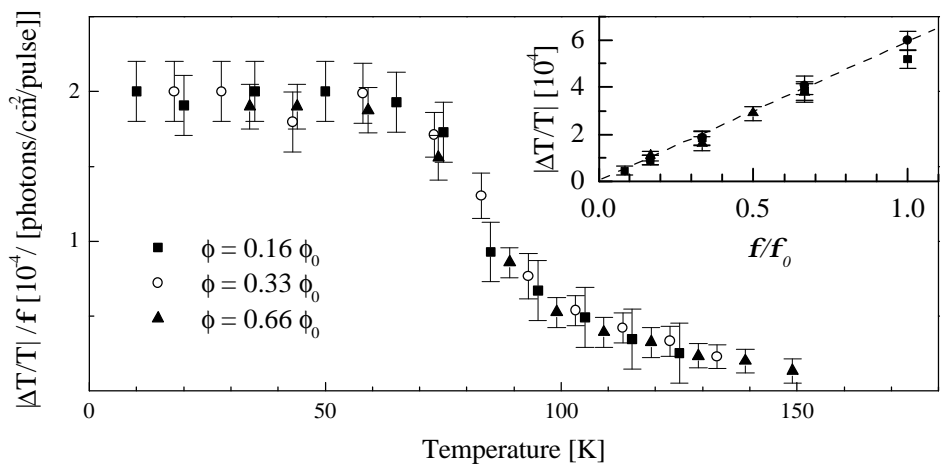

Figure 5: The T-dependence of the fast component amplitude, $|\Delta T/T|$, for a sample with $T_c = 89K$ using three different laser intensities. The data has been normalized with respect to the laser fluence, $\Phi$. The insert shows low temperature $|\Delta T/T|$ vs. $\Phi$, where $|\Delta T/T|$ scales linearly with $\Phi$. $\Phi_0 = 5.5 \; 10^{13}$ photons/cm$^2$/pulse.

**Time-resolved Photoinduced Spectroscopy: Processes, Model and Experiment**

**Initial Hot Carrier Relaxation.** An ultrashort laser pulse with an energy $E_L = 1\text{-}2$ eV incident on a HTS results in the creation of an electron and a hole with a relative energy $E_L$ (Figure 6a). For low and intermediate photoexcitation carrier densities (compared to the normal state carrier density), the process of energy relaxation can be described well by Allen's model (*22*). The model is applicable for intra-band relaxation in a metal or a semiconductor and has also been used successfully for determination of the electron-phonon coupling constant $\lambda$ in conventional metallic superconductors using time-resolved experiments performed at room temperatures (*23*).

The photoexcited carriers first termalize among themselves via intra-band scattering with a characteristic electron-electron (*e-e*) relaxation time of $t_{e-e} = \hbar E_F / 2pE^2 \approx 10 fs$, where $E$ is the carrier energy measured from the Fermi energy $E_F$. This quasiparticle avalanche multiplication due to electron-electron collisions takes place as long as $t_{e-e}$ is less then the electron-phonon relaxation time $t_{e-ph}$. Further energy relaxation occurs by electron-phonon scattering which occurs on a timescale given by the electron-phonon relaxation time (*22*) $t_{e-ph} = T_e/3\lambda\langle\omega^2\rangle$, where $T_e$ is initial carrier temperature, $T_e = E_I/C_e$, $E_I$ is the energy density per unit volume deposited by the laser pulse, $C_e$ is the electronic specific heat, and $\omega$ is the characteristic phonon frequency.

Electron-phonon relaxation time $t_{e-ph}$ and hence $\lambda$ has been determined experimentally for the case of YBa$_2$Cu$_3$O$_{7-\delta}$ (*14*), as well as Bi$_2$Sr$_2$CaCu$_2$O$_8$ and Bi$_2$Sr$_2$Ca$_2$Cu$_3$O$_{10}$ (*15*) from relaxation time fits in time-resolved experiments using intense laser pulses. For YBa$_2$Cu$_3$O$_{7-\delta}$ $t_{e-ph}$ has been found to be $t_{e-ph} = T_e/3\lambda\langle\omega^2\rangle \approx 100fs$ for initial carrier temperatures $T_e = E_I/C_e$ in the range 3000 K (*14*). A different group found $\tau_{e-ph} \approx 60fs$ for $T_e \approx 410$ K (*15*) giving $\lambda$ to be in the range $0.9 < \lambda < 1$. A similar $\lambda$ was also found in $Bi_2Sr_2CaCu_2O_8$ (*15*). Thus in the absence of a gap in the low-frequency spectrum, for the range of typical photoexcitation densities $E_I$ used in these experiments the relaxation process is complete within 10-100 fs. This is the usual case which occurs in metals (*23*).

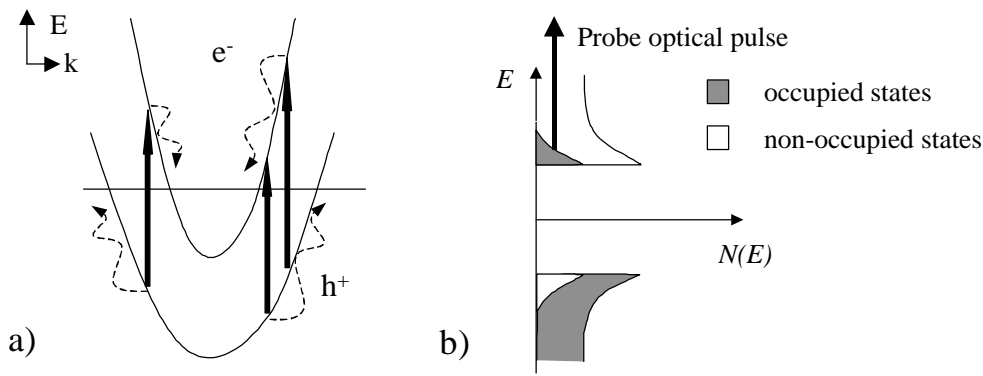

Figure 6: a) A scheme of the relaxation processes following the inter-band photoexcitation. If there is no gap (or "pseudogap") in the density of states near $E_F$ the equilibrium at some slightly elevated lattice temperature would be achieved within 10-100fs. b) When the gap in the density of states is present, it creates the bottleneck for the quasiparticle relaxation. The photoexcited quasiparticles together with high-frequency phonons form a near steady state distribution with the relaxation time governed by anharmonic phonon decay (*9*). The variably delayed probe pulse therefore measures the temporal change in the photoexcited carrier density.

In a superconductor, however, the presence of a superconducting gap creates a relaxation bottleneck. The carriers therefore accumulate in quasiparticle states above the gap waiting to recombine into pairs. Their typical recombination time $\tau_R$ is 1-2 orders of magnitude longer than the initial intra-band Allen relaxation process (*9*) and can be probed by excited state absorption using a second probe laser pulse (Figure 6b).

It is the existence of this near-equilibrium QP population that gives us a powerful tool to investigate different aspects of the gap structure in high-$T_c$ superconductors. Because the recombination dynamics depends on the density of electronic states at low energies, the time-resolved experiments give detailed information about the *T*-dependence, doping dependence and symmetry of the gap.

**The Laser Probe Process.** For probing the QP states it is quite important that the energy of the probe laser photons is above the plasma frequency of the material, because in this case we can make the simple assumption, that the probe absorption is due to an unscreened inter-band transition between the QP states and a band well above $E_F$ (Figure 6b). The absorbance can then be approximated by the Fermi golden rule.

The *change* in the absorption probability is then given by $\delta r \propto \delta(-n_i \rho_f |M_{ij}|^2) \approx -\delta n_{pe} (\rho_f |M_{ij}|^2)$, where $n_{pe}$ is the photoexcited quasiparticle density, $\rho_f$ is the final

                                                                                                                                       10

density of unoccupied states and $M_{ij}=\langle\mathbf{p}\cdot\mathbf{A}\rangle$ is the dipole matrix element. The simplest case we can consider is to assume that the matrix element $M_{ij}$ and the density of final states $\rho_f$ are not changed by the photoexcitation process (the adiabatic approximation). Then the amplitude of the photoinduced absorption, $|DA/A|$, is proportional to the photoinduced transmission amplitude, $-|DT/T|$ (and in the linear approximation also to $|\Delta R/R|$), which is in turn proportional to the change in the number of photoexcited quasiparticles $|DA/A| \propto -|DT/T| \propto \delta r \approx -\delta n_{pe}(\rho_f |M_{ij}|^2)$.

In a complete description, the effects of the change in final states $\rho_f$ and the matrix element $M_{ij}$ (described by the Frank-Condon term $-n_i \delta(\rho_f |M_{ij}|^2)$) by the photoexcited particles need to be considered as well, since they can be important contributions, especially if polaronic lattice distortions occur around them. However, time-resolved spectroscopy enables these effects to be separated because they have very different dynamics (longer lifetimes).

When the Frank-Condon term is neglected, the probe signal is simply weighted by $\rho_f |M_{ij}|^2$. In the case of cuprates when $E_{probe} \approx 1.5$ eV, this corresponds closely to the O-Cu charge-transfer dipole transition, so the final state band in the probe process can also be identified (*8*).

**Theoretical Model for the Temperature Dependence of the Photoinduced Reflection or Transmission.** A theoretical description of the temperature dependence of the photoinduced transmission or reflection amplitude involves a calculation of the photoexcited carrier density $n_{pe}$ as a function of temperature. Such a calculation was performed recently by Kabanov et al. (*9*).

The theoretical model for the photoinduced transmission amplitude, $|DT/T|$, predicts very different temperature dependences, depending on the temperature dependence of the gap itself. In the case of a temperature-independent gap,

$$\left|\Delta T/T\right| \propto -\frac{1}{\Delta_0}\left[1 + \frac{2\nu}{N(0)\hbar\Omega} e^{-\Delta_0/k_B T}\right]^{-1}, \qquad (4)$$

where $\Delta_0$ is the energy gap, $\nu$ is the effective number of phonons per unit cell emitted in the QP recombination process, $\Omega$ is the typical phonon frequency and $N(0)$ is the density of states at $E_F$. The calculated T-dependence of the photoinduced transmission is shown in Figure 7. The predicted photoinduced transmission amplitude falls to zero asymptotically at high temperatures and one cannot easily identify a *temperature* associated with such a T-independent gap.

On the other hand, if the gap closes at a well defined temperature due to a collective effect as in the BCS scenario, such that $\Delta(T) \to 0$ as $T \to T_c$, the relaxation bottleneck clearly disappears at this temperature and the formula for the temperature dependence of the photoinduced signal amplitude is somewhat modified:

$$\left|\Delta T/T\right| \propto -\frac{1}{\Delta(T)+k_B T/2}\left[1 + \frac{2\nu}{N(0)\hbar\Omega}\sqrt{2k_B T/p\Delta(T)} \cdot e^{-\Delta(T)/k_B T}\right]^{-1}, \qquad (5)$$

such that the photo-induced transmission amplitude drops to zero at $T_c$.

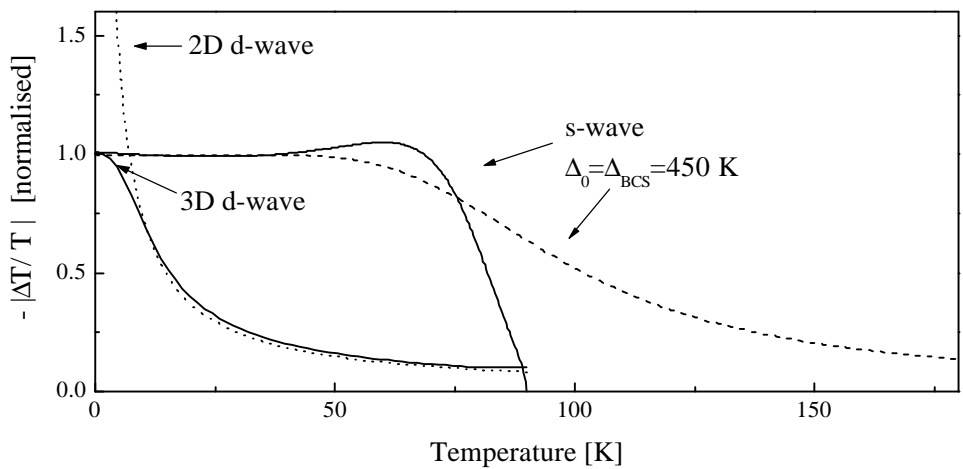

Figure 7: Theoretical predictions (*9*) for the amplitude of the photoinduced transmission as a function of temperature for the cases of temperature dependent BCS-like gap ("knee"-shaped upper solid line) and temperature independent s-wave gap (dashed line) with the same gap values, and for the 2D and 3D d-wave gap (dotted and solid line respectively).

It should be emphasized that the two cases are qualitatively different and can easily be experimentally distinguished. A T-dependent BCS-like collective gap, which closes at $T_c$, cannot be used to describe the asymptotic behavior at high temperatures, while in the T-independent gap model $|\Delta T/T|$ falls far too slowly at high T to be able to describe the data near optimum doping. Furthermore, in the case of a BCS-like T-dependent gap, there is also strong temperature dependence of the relaxation time which diverges close to $T_c$ as $\tau_R \propto 1/\Delta(T)$ (*3,8,9*) since $\Delta(T) \to 0$ as $T \to T_c$ which is absent in the case of T-independent gap (see insert to Figure 3).

In the model calculations so far we considered the case of a T-independent gap and a BCS-like T-dependent gap. In both cases the gap was considered to be isotropic. We finish this section by considering the effect of a strongly anisotropy gap, applicable for example for the case of a pure *d*-wave superconductor. The effect of the finite density of states at low energies is that there is no real bottleneck in the QP relaxation and as a consequence we expect an increase in the photoinduced transmission only at the lowest temperatures, well below $T_c$. The calculated photoinduced transmission as a function of T is shown in Figure 4 for the case of a 2-dimensional DOS and a 3-dimensional d-wave DOS.

**Comparison With the Data on $YBa_2Cu_3O_{7-\delta}$.** The temperature dependence of photoinduced transmission and reflection for a large number of different oxygen concentrations have shown (*8,9*) that the behavior in underdoped samples with $\delta > 0.15$ is quite different than near optimum doping where $\delta < 0.15$. The data for three different samples with different doping $\delta$ are plotted as a function of temperature in Figure 8.

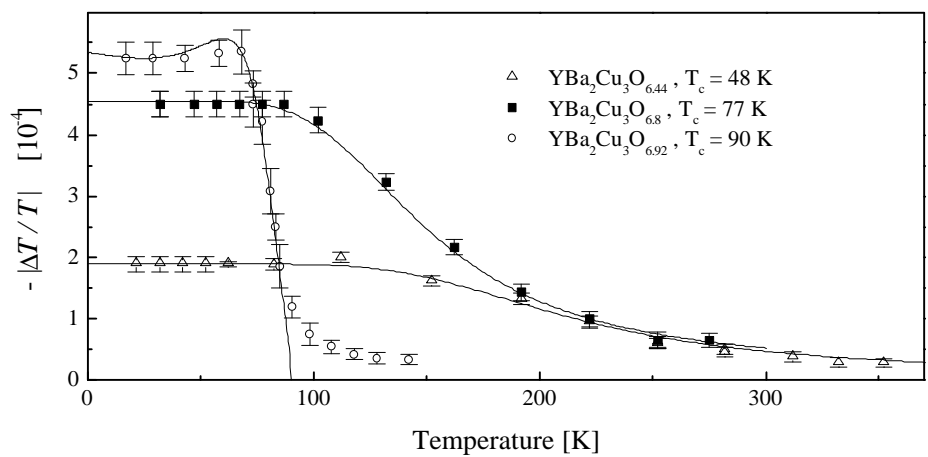

Figure 8: The temperature dependence of the photoinduced transmission amplitude for near-optimally doped sample and two underdoped samples. The solid lines are the fits to the data using equation 5 with a T-dependent BCS-like gap for near-optimally doped sample and the equation 4 with the T-independent pseudogap for the underdoped samples.

In the underdoped samples the photoinduced transmission amplitude, $|DT/T|$, drops rather slowly with increasing temperature, while the data near optimum doping show an abrupt drop in the $|DT/T|$, falling to near-zero close to $T_c$. Moreover, near optimum doping, a slight maximum is clearly observed in the data just below $T_c$ as predicted by theoretical model (Figure 7). The calculated temperature-dependence of the photoinduced transmission is superimposed on the data in Figure 8. For $\delta > 0.15$, a T-independent gap fit was used, while for $\delta < 0.15$ the data is fitted by equation 5 for a T-dependent BCS gap, in both cases with excellent agreement with the data. The data on photoinduced transmission amplitude together with the relaxation time data thus provide clear evidence for the existence of a T-independent gap in the underdoped $YBa_2Cu_3O_{7-\delta}$ and a cross-over to a BCS-like gap near optimum doping. The temperature-dependence of the photoinduced transmission is actually a function of $\Delta/T$ in both cases, which means that the T-dependences of different samples can be scaled onto a single curve (9), provided the correct T-dependence of the gap is used. This enables us to define a criterion by which a temperature scale can be associated with the gap. In the BCS case when the gap closes at $T_c$, this scale is $T_c$ itself. However, in the case of underdoped samples, where the fall-off at high temperatures is asymptotic, we chose the onset of a "pseudogap" at $T^*$ where the amplitude of the signal falls to 10% of maximum. The main reason for choosing the 10% point and not the mid-point for example is to enable comparison with the literature where the "onset" is the common criterion. A plot of the pseudogap temperature as a function of doping obtained in this way is plotted in Figure 9.

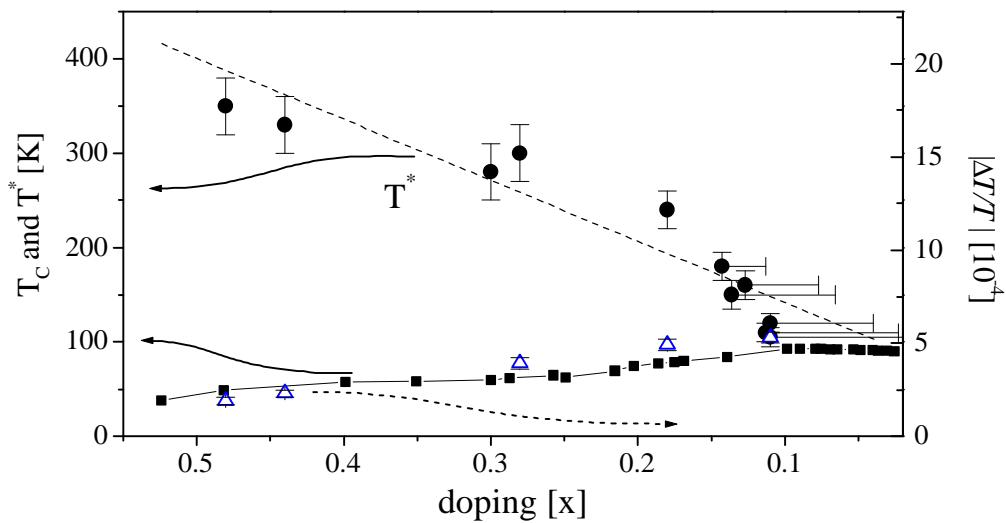

Figure 9: The pseudogap onset temperature (solid circles) for $YBa_2Cu_3O_{7-x}$ determined from time-resolved measurements together with the $DT/T$ amplitude taken at series of samples of the same thickness and the same substrate in single experimental run.

Also shown in the plot is the magnitude of the photoinduced transmission signal at low temperature (T << $T_c$) as a function of doping $x$, which increases linearly with increasing hole concentration $n_h \propto (1 - x)$ in the $CuO_2$ planes. From equation 4, the amplitude of the photoinduced transmission is predicted to vary inversely with the magnitude of the $DT/T$. From the linear relationship between $|DT/T|$ and $x$ one can deduce that the gap is inversely proportional to doping $\Delta_0 \propto 1/n_h$ in a large part of the phase diagram (*24*).

It is worth mentioning here that the cross-over between a T-independent gap and a T-dependent gap near optimum doping does not appear to be well defined, but there appears to be a small, but clearly visible non-zero signal in the $|DT/T|$ above $T_c$ (Figure 8). This is believed to be a result of the inhomogeneity of the hole density in the $CuO_2$ planes near optimum doping, where there appears to be a coexistence of regions with a collective gap and regions with a "local" T-independent gap. This observation would appear to be consistent with the two-component paradigm near optimum doping in $YBa_2Cu_3O_{7-\delta}$ (*25*).

**Conclusion**

In the present chapter we have described the time-resolved photoinduced absorption spectroscopy and its application in investigation of the low energy structure in high-$T_c$ superconductors. Because time-domain spectroscopy enables the separation of different spectral components in the low energy excitation spectrum by their lifetime, it can therefore resolve long-standing controversial issues regarding the origin of the different components in the low-energy spectra of the cuprates, which inevitably arise in the interpretation of frequency-domain spectroscopy data. This applies particularly to separation of quasiparticle states from localized states.

The recent theoretical model (*9*), has been used to explain the temperature dependences of photoinduced transmission or reflection in $YBa_2Cu_3O_{7-\delta}$. In spite of the fact that the model makes some simplifications, for example regarding the anisotropy of the gap and does not consider a realistic phonon density of states, the agreement with experiment appears to be sufficiently good to justify these simplifications.

We have presented the results of the measurements of the T-dependence of the photoinduced transmission over wide range of doping in $YBa_2Cu_3O_{7-\delta}$. The data have given the first direct experimental evidence for a cross-over from a T-independent gap in underdoped state to a T-dependent gap near optimal doping in the cuprates. Complementary data on the QP recombination time $\tau_R$ (*3,8,9*), which was not discussed in detail here, give additional support to the conclusions regarding the existence of a T-independent pseudogap and T-dependent collective gap made here on the basis of photoinduced transmission amplitude analysis. Particularly important is the observation of a divergence of $\tau_R$ at $T_c$ in optimally doped $YBa_2Cu_3O_{7-\delta}$, which is a result of the fact that $\tau_R \sim 1/\Delta(T)$ (*9*) where $\Delta(T) \to 0$ at $T_c$. In the cross-over region the two gaps appear to coexist, suggesting a spatially inhomogeneous ground state configuration, in agreement with two-component models of the cuprates near optimum doping (*25,26*).

We conclude that the time-domain data on QP dynamics gives complementary information on the low-energy electronic structure of these materials, which cannot be obtained by more usual spectroscopic means and that the experimental method in conjunction with newly-developed theoretical models (*9*) is sufficiently general to be applicable also in other materials exhibiting gap in the low-energy excitation spectrum, such as charge-density wave systems or confined structures exhibiting a gap structure due to quantum size effects.